\documentclass[secnumarabic,amssymb, nobibnotes, aps, prd]{revtex4-2}
\usepackage{graphicx}
\usepackage{amsmath}
\usepackage{subcaption}

\usepackage{multirow}

\begin{document}
	\title{Revisiting thermodynamic topology of Hawking-Page and Davies type phase transitions}
	\author{Bidyut Hazarika$^1$}
	
	\email{$rs_bidyuthazarika@dibru.ac.in$}
	\author{Naba Jyoti Gogoi$^1$}
	\email{$rs_nabajyotigogoi@dibru.ac.in$}
	\author{Prabwal Phukon$^{1,2}$}
	\email{prabwal@dibru.ac.in}	
	\affiliation{$1.$Department of Physics, Dibrugarh University, Dibrugarh, Assam,786004.\\$2.$Theoretical Physics Division, Centre for Atmospheric Studies, Dibrugarh University, Dibrugarh, Assam,786004.}
	\begin{abstract}
		In this work, we propose a common vector field to study the thermodynamic topology of the Davies type and Hawking-Page phase transitions. Existing literature has shown that studying these two types of phase transitions typically requires defining two separate vector fields . In our approach, we adopt Duan's $\phi$-mapping topological current theory to define a novel vector field, denoted as $\phi$, whose critical points exactly correspond to the Davies point and the Hawking-Page phase transition point. More importantly, we can differentiate between these two points by their topological charge. While, the topological charge for the critical point corresponding to the Davies-type phase transition is found to be $-1$, the same for the Hawking-Page phase transition point, it is $+1$. Although our analysis is applicable to all black hole systems where both types of phase transitions are found, we illustrate it using three simple systems as examples: the Schwarzschild AdS black hole, the Reissner-Nordström AdS black hole in the grand canonical ensemble, and finally the Kerr AdS black holes in the grand canonical ensemble. It is wellknown that these black holes exhibit both Davies and Hawking-Page phase transitions. With our proposed vector $\phi$, the critical points obtained for these three systems exactly match the Davies-type and Hawking-Page phase transition points, and the associated topological charges are found to be  $-1$ for the Davies point and $+1$ for the Hawking-Page phase transition point.
	\end{abstract}
	
	\maketitle
	
	
	
	
	
\section{introduction}

In the early 1970s, the groundwork in black hole thermodynamics established the basis for comprehending the correlation between black holes and the laws of thermodynamics \cite{Bekenstein:1973ur,Hawking:1974rv,Hawking:1975vcx,Bardeen:1973gs}. Subsequent research endeavors have unveiled numerous compelling findings in this field \cite{Wald:1979zz,bekenstein1980black,Wald:1999vt,Carlip:2014pma,Wall:2018ydq,Candelas:1977zz,Mahapatra:2011si}.
One very interesting aspect in the study of  black hole thermodynamics is the issue phase transition\cite{Davies:1989ey,Hawking:1982dh,curir_rotating_1981,Curir1981,Pavon:1988in,Pavon:1991kh,OKaburaki,Cai:1996df,Cai:1998ep,Wei:2009zzf,Bhattacharya:2019awq,Kastor:2009wy,Dolan:2010ha,Dolan:2011xt,Dolan:2011jm,Dolan:2012jh,Kubiznak:2012wp,Kubiznak:2016qmn,Bhattacharya:2017nru}. Black holes undergo various types of phase transitions, including Davies type phase transition\cite{Davies:1989ey},  Hawking–Page phase transition \cite{Hawking:1982dh}, extremal phase transition ( transition of black holes from non-extremal to extremal states) \cite{curir_rotating_1981,Curir1981,Pavon:1988in,Pavon:1991kh,OKaburaki,Cai:1996df,Cai:1998ep,Wei:2009zzf,Bhattacharya:2019awq} and van der Waals-type phase transition (a phase transition reminiscent of van der Waals behavior)\cite{Kastor:2009wy,Dolan:2010ha,Dolan:2011xt,Dolan:2011jm,Dolan:2012jh,Kubiznak:2012wp,Kubiznak:2016qmn,Bhattacharya:2017nru}. 
Phase transitions of black holes have been extensively studied using different frameworks such as extended black hole thermodynamics \cite{14,15,16,17,18,19,20,21,22,23,24,25,new}, restricted phase space thermodynamics and holographic thermodynamics \cite{rp1,rp2,rp3,rp4,rp5,rp6,rp7,rp8,rp9,rp10,rp11}, geometrical interpretation of phase transition \cite{Quevedo:2008ry,Akbar:2011qw,Hendi:2015cka,Sarkar:2006tg,Hendi:2015xya,Banerjee:2016nse,Bhattacharya:2017hfj,Bhattacharya:2019qxe,Gogoi:2023qni,Kumar:2012ve} etc. A relatively recent development in this direction is the topological interpretation of black hole phase transition \cite{Wei:2022dzw,Wei:2021vdx,Yerra:2022alz,Yerra:2022eov,Gogoi:2023xzy,Gogoi:2023qku,Gogoi:2023wih,Yerra:2022coh,Yerra:2023ocu,Barzi:2023msl,Ahmed:2022kyv,Wei:2022mzv,Fan:2022bsq,Wu:2022whe,Fang:2022rsb,Wu:2023xpq,Wu:2023sue,Li:2023ppc,Wei:2023bgp,Alipour:2023uzo,Zhang:2023uay,Sadeghi:2023aii,Wang:2024zbp,Shahzad:2024ojx,Malik:2024kau,Zhao:2024tlu,Wu:2024rmv,Hazarika:2024cpg,Hazarika:2023iwp,Sadeghi:2024krq,Zhang:2023svu,Bhattacharya:2024bjp} in which the critical points are endowed with some topological charges.  Some of these works have used  thermodynamic topology to understand the van der Waals-type phase transition \cite{Wei:2021vdx,Yerra:2022alz,Yerra:2022eov,Gogoi:2023xzy,Gogoi:2023qku} , the Hawking-Page phase transition \cite{Yerra:2022coh,Yerra:2023ocu,Barzi:2023msl,yerrabm} and Davies-type phase \cite{Bhattacharya:2024bjp}.  These existing literatures suggest that topologically investigating Hawking-Page and Davies-type phase transition typically requires defining two separate potentials or vector fields \cite{Zhang:2023uay,Yerra:2022coh,Bhattacharya:2024bjp,Fan:2022bsq}.
The key inspiration behind thermodynamic topology of black holes lies in Duan's $\phi$-mapping current theory\cite{Duan,Duan:2018rbd}. A simple and brief discussion of this approach is provided below:

In Duan's method, for a vector field $\phi=\{\phi^a\}$ (with $a=1,2$) in the coordinate space $x^\nu=\{t,r_+,\theta\}$, one can construct a topological current 
\begin{equation}
	j^\mu=\frac{1}{2\pi}\epsilon^{\mu \nu \rho}\epsilon_{ab}\partial_\nu n^a\partial_\rho n^b,
\end{equation}
where, $\partial_\nu=\frac{\partial}{\partial x^\nu}$ and $\mu,\nu,\rho=0,1,2$. Also, $n^a$ is the normalized vector defined as 
\begin{equation}
	n^a=\frac{\phi^a}{||\phi||}; \quad a=1,2  \quad \text{with} \quad \phi^1=\phi^{r_+}, \quad \phi^2=\phi^{\theta}.
\end{equation}
The normalized vector $n^a$ satisfies the condition
\begin{equation}
	n^an^a=1 \quad \text{and} \quad n^a\partial_\nu n^a=0.
\end{equation}
Through simple analysis, it is easy to check for the conservation of the topological current
\begin{equation}
	\partial_\mu j^\mu=0.
\end{equation}
Using the Jacobi tensor $\epsilon^{ab} J^\mu \left(\frac{\phi}{x} \right)=\epsilon^{\mu\nu\rho} \partial_\nu \phi^a \partial_\rho \phi^b$ with the two-dimensional  Laplacian Green function $\Delta_{\phi^a}\ln ||\phi||=2\pi \delta^2(\phi)$, the topological current can be expressed as \cite{Duan}
\begin{equation}
	j^\mu=\delta^2(\phi)J^\mu\left(\frac{\phi}{x}\right).
\end{equation}
This suggest that $j^\mu$ is non-zero only at the point where $\phi^a(x^i)=0$ where, the $i$-th solution is denoted as $\vec{x}=\vec{z}_i$. Using the $\delta$-function theory \cite{Schouten1951}, the density of topological current can be obtained as  
\begin{equation}
	j^0=\sum_{i=1}^N \beta_i\eta_i\delta^2 (\vec{x}-\vec{z}_i).
\end{equation}
Here, $\beta_i$ is the Hopf index and its positive value quantifies the number of loops formed by $\phi^a$ in the vector $\phi$ space as $x^\mu$ traverses around the zero point $z_i$. The Brouwer degree $\eta_i$ is given as $\eta_i =\text{sign}\left( J^0(\phi/x)_{z_i}\right)=\pm1$. Then, within a parameter region $\Sigma$, the corresponding topological charge can be calculated as 
\begin{equation}
	Q=\int_\Sigma j^0d^2x=\sum_{i=1}^N \beta_i\eta_i=\sum_{i=1}^N w_i,
\end{equation}
were $w_i$ represents the winding number for $i$-th zero point of $\phi$.\\

In this work, we introduce a new potential, denoted as $\phi$, in Section II, which serves as a common potential for studying both the Hawking-Page and Davies-type phase transitions. Our objective is to construct a vector field in such a manner that its zero points, or critical points, correspond to the Davies point and the Hawking-Page phase transition point. We provide examples of black holes wherein both types of phase transitions coexist, namely the Schwarzschild AdS, RN AdS in the grand canonical ensemble, and Kerr AdS black holes in the grand canonical ensemble.\\

Our paper is organized as follows: In section \textbf{$II$}, we have introduced a new vector field utilizing which topology of Hawking-Page transition and Davies-type phase transition. can be studied, in section \textbf{$II$},\textbf{$III$},\textbf{$IV$} we have studied these two types of phase transition for Schwarzschild, Reissner-Nordstr\"{o}m and Kerr black holes in anti-de Sitter space respectively. In section \textbf{$V$}, we have presented our concluding remarks.
\section{New Vector Field}
We propose a new two-dimensional vector field to simplify the study of the topology of Hawking-Page transition and Davies-type phase transition. We aim to construct a vector field such that its critical points or zero points exactly match with the Hawking-Page (HP) transition point and Davies point. The components of the proposed vector field are given by :
\begin{eqnarray}
	\begin{cases}
		\phi=\left(\phi^S,\phi^\theta \right)=\left( \frac{1}{S} \frac{\partial F^2}{\partial S},-\cot\theta ~\csc\theta  \right)\\
		\phi=\left(\phi^r,\phi^\theta \right)=\left( \frac{1}{r_+} \frac{\partial F^2}{\partial r_+},-\cot\theta ~\csc\theta  \right)
	\end{cases}
	\label{newfield}
\end{eqnarray}
Depending on whether the thermodynamic quantities are expressed in terms of entropy $S$ or horizon radius $r_+$, the components will be defined by the first or the second equation stated above. In these expressions,  $F$ is the free energy. The basic idea behind this proposed vector field is explained below :
Taking the first component of the vector field :
\begin{equation}
	\phi^S= \frac{1}{S}\frac{\partial F^2}{\partial S}=\frac{2 F}{S} \frac{\partial F}{\partial S}=-2 F \frac{\partial T}{\partial S}=-2 F \frac{T}{C}
\end{equation}

Where, we have used $F=M-TS$ and $T=\frac{\partial M}{\partial S}$ and $C$ is the specific heat. Therefore, by construction, the zero points of the vector field given by $\phi^S=0$ implies
\begin{equation}
	F=0 \hspace{0.5 cm}\text{and} \hspace{0.5 cm} \frac{1}{C}=0
	\label{zeropoints}
\end{equation}

While $F=0$ is the point where Hawking-Page phase transition occurs and $\frac{1}{C}=0$ is the Davies point. Hence it can be inferred that the zero points of the vector field in eq.\ref{newfield} are exactly the Hawking-Page and Davies type phase transition points. A similar argument can be made in case of the second vector function written in terms of $r_+$. To demonstrate the equivelence of these two expressions in $S$ and $r_+$, we perform our analyses in $r_+$ variable for Schwarzchild AdS and RN AdS black hole cases and in $S$ variable in Kerr AdS black hole case.

	\section{Schwarzchild Ads Black hole}
\begin{figure}[h]	
		\centering
		\includegraphics[width=0.4\linewidth]{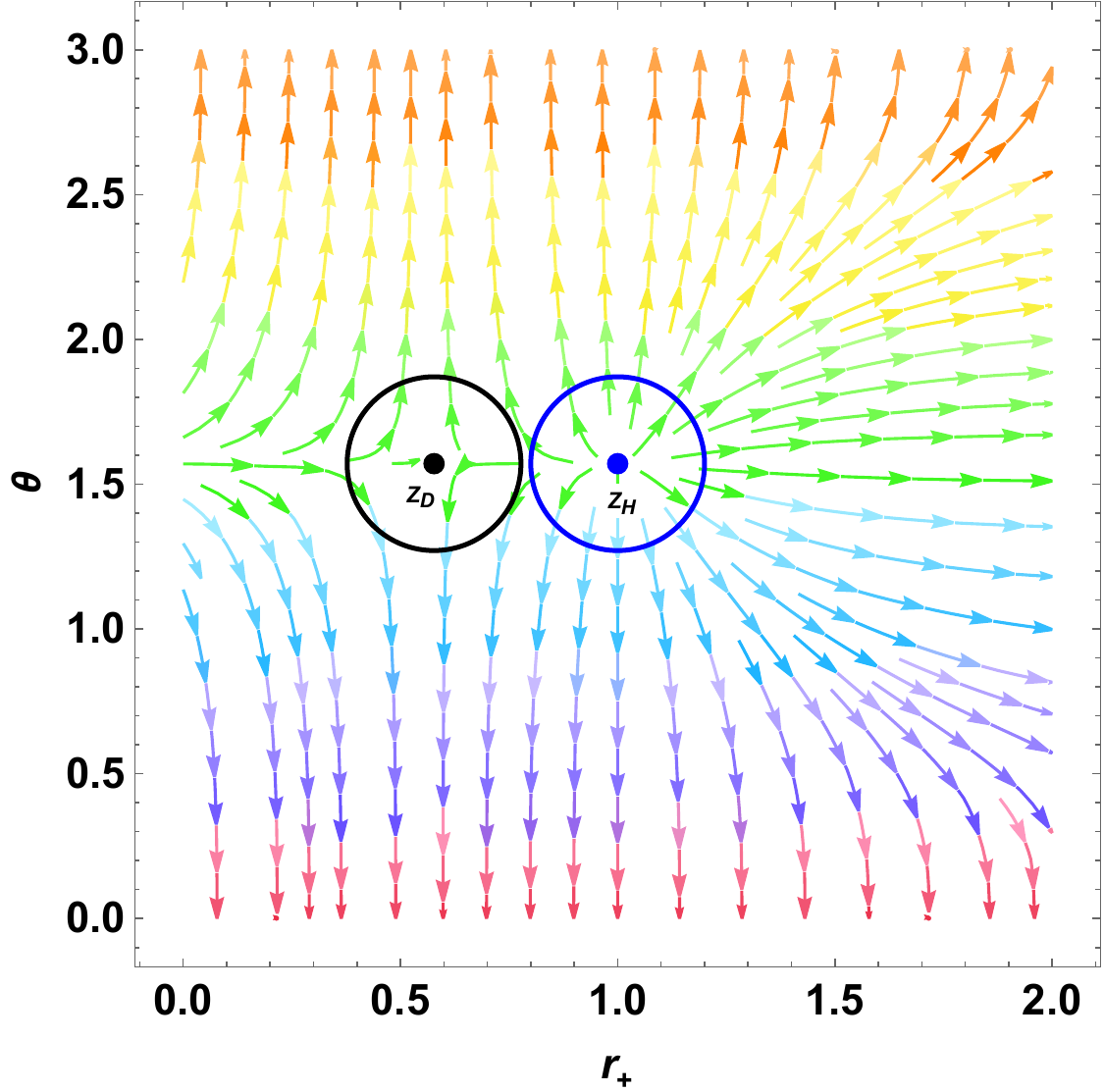}
		\caption{  Vector plot of a portion of the field ($n^1,n^2$) for Schwarzchild AdS black hole. $Z_D$ and $Z_H$ represent the Davies point and Hawking-Page phase transition point respectively.
		}
		\label{2}
	\end{figure}

\begin{figure}[h]	
		\centering
		\begin{subfigure}{0.40\textwidth}
			\includegraphics[width=\linewidth]{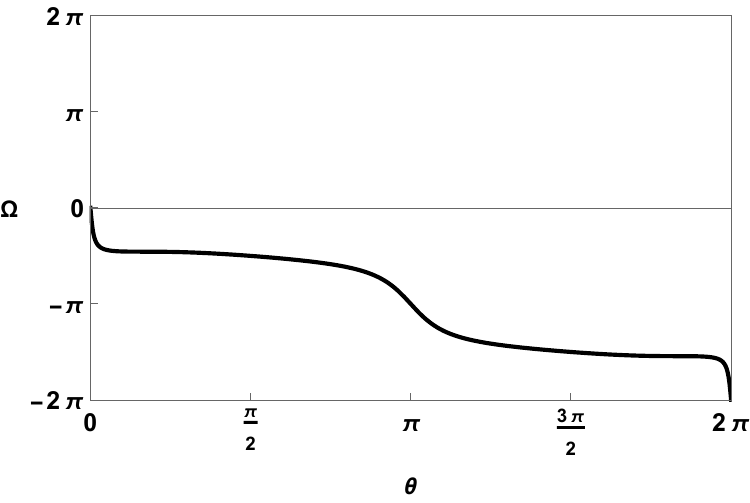}
			\caption{}
			\label{3a}
		\end{subfigure}
		\hspace{0.5cm}
		\begin{subfigure}{0.40\textwidth}
			\includegraphics[width=\linewidth]{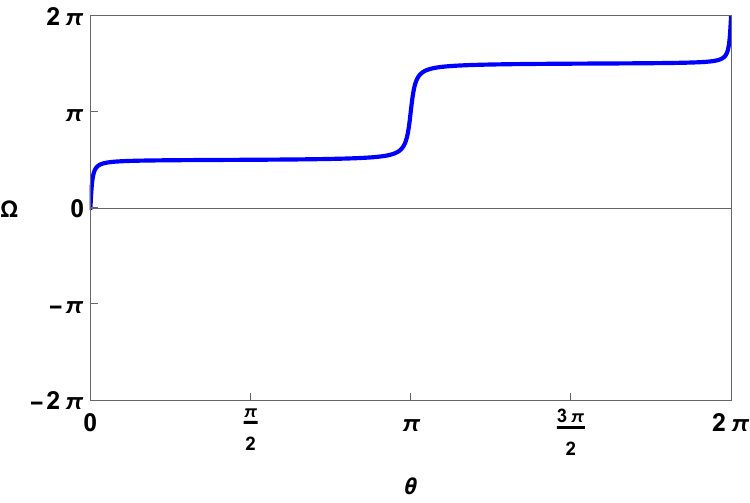}
			\caption{}
			\label{3b}
		\end{subfigure}
		
		\caption{Schwarzchild AdS black hole: FIG.(a) represents topological number calculation  for Davies point around the black contour in FIG.(\ref{2}), and FIG.(b) represents the topological number calculation  for Hawking-Page phase transition point around the blue contour in FIG.(\ref{2}),
		}
		\label{3}
	\end{figure}

	The metric for Schwarzchild-Ads is given by 
	\begin{equation}
		ds^2=-\left(1 - \frac{2 M}{r} + \frac{r^2}{b^2}\right) dt^2+\left(1 - \frac{2 M}{r} + \frac{r^2}{b^2}\right)^{-1} dr^2+r^2 d\Omega
	\end{equation}
	where b is the curvature radius of the Ads spacetime and can be written in terms of pressure as follows 
	$$P=\frac{3}{8 \pi b^2}$$
	For simplicity, we set $b=1$ and find out the mass as follows 
	\begin{equation}
		M=\frac{1}{2} \left(r_+^3+r_+\right)
		\label{mass schw}
	\end{equation}
	The temperature is calculated by differentiating the mass with respect to entropy as follows.
	\begin{equation}
		T=\frac{d M}{d S}=\frac{3 r_+^2+1}{4 \pi  r_+}
		\label{temp schw}
	\end{equation}
	Using equations eq.(\ref{mass schw}) and eq.(\ref{temp schw}), We calculate the free energy as :
	\begin{equation}
		F=M-TS=\frac{1}{4} \left(r_+-r_+^3\right)
		\label{free schw}
	\end{equation}
	\begin{figure}[h]	
		\centering
		\begin{subfigure}{0.40\textwidth}
			\includegraphics[width=\linewidth]{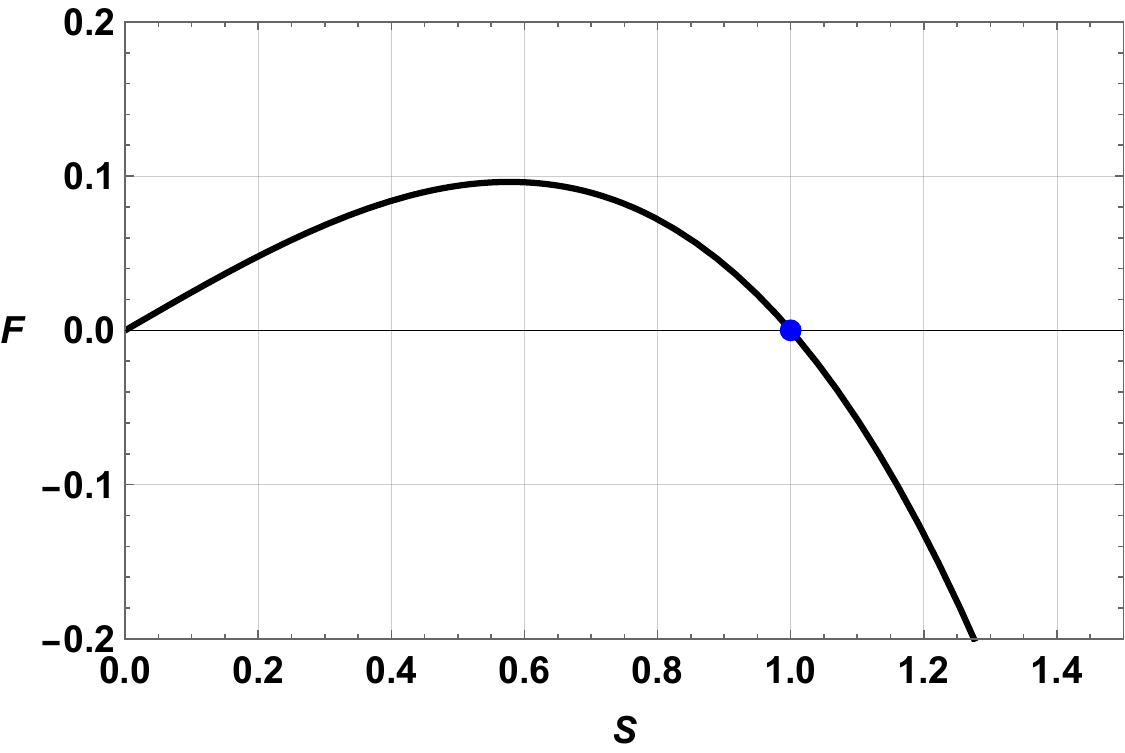}
			\caption{}
			\label{1a}
		\end{subfigure}
		\hspace{0.5cm}
		\begin{subfigure}{0.40\textwidth}
			\includegraphics[width=\linewidth]{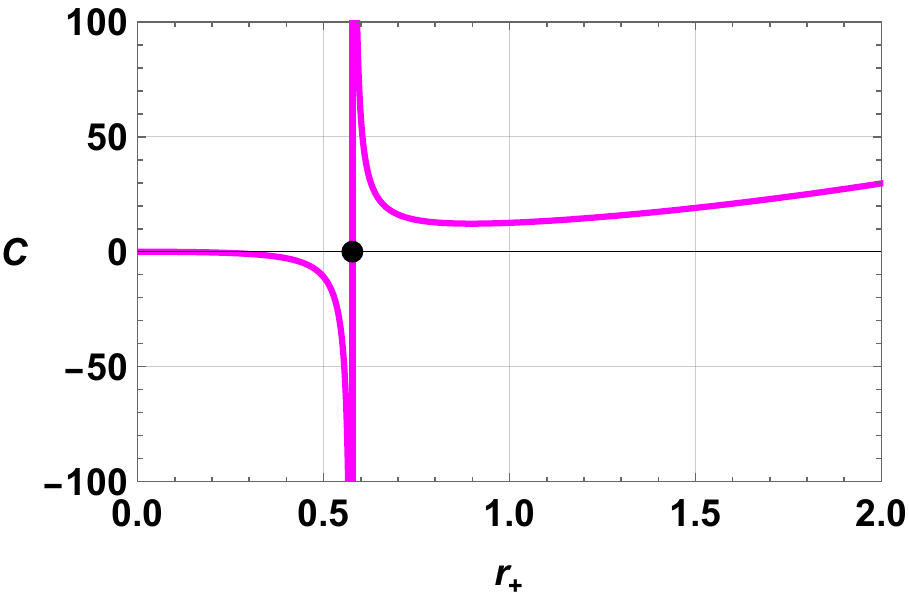}
			\caption{}
			\label{1b}
		\end{subfigure}
		
		\caption{Figure (a) represents $F$ vs $r_+$ plot for Schwarzchild AdS black hole. The blue dot represents $r_+=1$, the free energy becomes zero. This is the exact point at which the Hawking-Page transition takes place. FIG.(b) represents $C$ vs $r_+$ plot. The black dot represents at $r_+=\frac{1}{\sqrt{3}}$, the specific heat diverges. This is the exact point at which Davies-type transition takes place
		}
		\label{1}
	\end{figure}
	The exact point at which the Hawking-Page transition takes place is found by setting $F=0$. As shown in the FiIG.(\ref{1a},)1 at $r_+=1$, Hwking-Page transition takes place.\\ 
	The specific heat of the system is:
	\begin{equation}
		C=T\frac{\partial S}{\partial T}=\frac{2 \pi  r_+^2 \left(3 r_+^2+1\right)}{3 r_+^2-1}
	\end{equation}
	at $S=\frac{1}{\sqrt{3}}$ the specific heat diverges. This is the exact point at which Davies type transition takes place as shown in FIG.(\ref{1b}).\\
	Now applying equations eq.(\ref{free schw}) in eq. (\ref{newfield}), we get the component as :
	\begin{equation}
		\phi^r_+=\frac{1}{8} \left(3 r_+^4-4 r_+^2+1\right)
	\end{equation}
	and 
	\begin{equation}
		\phi^\theta=-\cot (\theta ) \csc (\theta )
	\end{equation}
	Using the components the unit vectors are calculated as 
	\begin{equation}
		\begin{cases}
			n^1=\frac{3 r_+^4-4 r_+^2+1}{8 \sqrt{\cot ^2(\theta ) \csc ^2(\theta )+\frac{1}{64} \left(3 r_+^4-4 r_+^2+1\right){}^2}}\\
			n^2=-\frac{\cot (\theta ) \csc (\theta )}{\sqrt{\cot ^2(\theta ) \csc ^2(\theta )+\frac{1}{64} \left(3 r_+^4-4 r_+^2+1\right){}^2}}
		\end{cases}
	\end{equation}
	We plot a portion of the vector field ($n^1,n^2$) in FIG.(\ref{2}). As we can see from FIG.(\ref{2}),  there are two critical points or zero points in that particular portion of the vector field ($n^1,n^2$). One is at $Z_D=0.57735$ and the other at $Z_H=1$. Comparing with FIG.(\ref{1}),, we can see $Z_D$(enclosed by the black contour) is the Davies point at which Davies type phase transition occurs and $Z_H$ is the point(enclosed by the blue contour) at which Hawking-Page phase transition occurs. \\
	To calculate the topological charge, we perform a contour integral around the black and the blue colored contour. In FIG.(\ref{3a},) as represented by the black solid line,  the topological charge corresponds to Davis point is found to be $-1$ and the topological charge for Hawking-Page phase transition is $1$ as represented by the blue solid line in FIG.(\ref{3b}).
	\section{RN AdS black hole in grand canonical ensemble}
\begin{figure}[h]	
		\centering
		\begin{subfigure}{0.4\textwidth}
			\includegraphics[width=\linewidth]{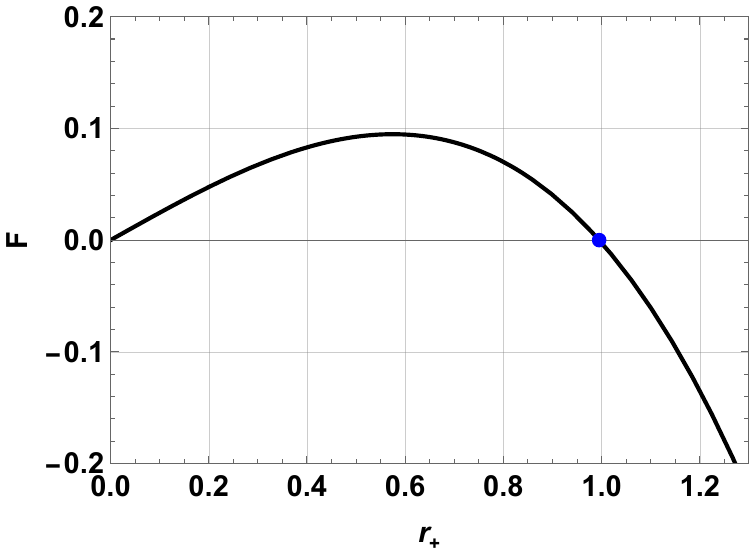}
			\caption{}
			\label{4a}
		\end{subfigure}
		\hspace{0.5cm}
		\begin{subfigure}{0.4\textwidth}
			\includegraphics[width=\linewidth]{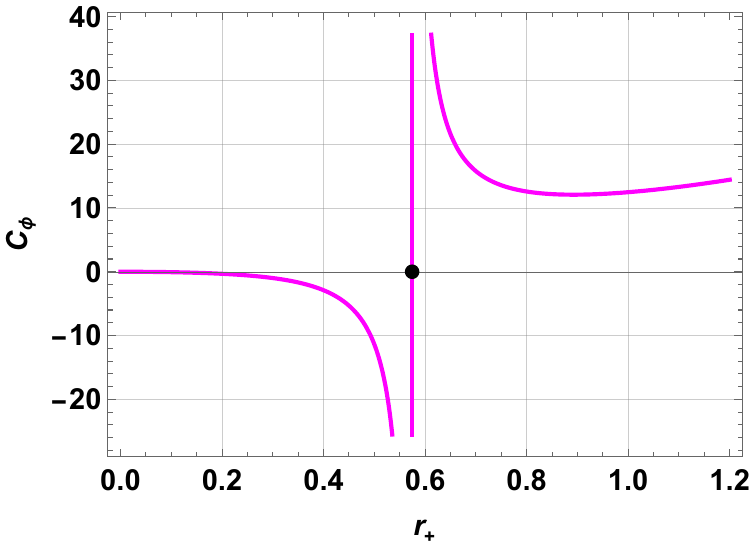}
			\caption{}
			\label{4b}
		\end{subfigure}
		\begin{subfigure}{0.40\textwidth}
			\includegraphics[width=\linewidth]{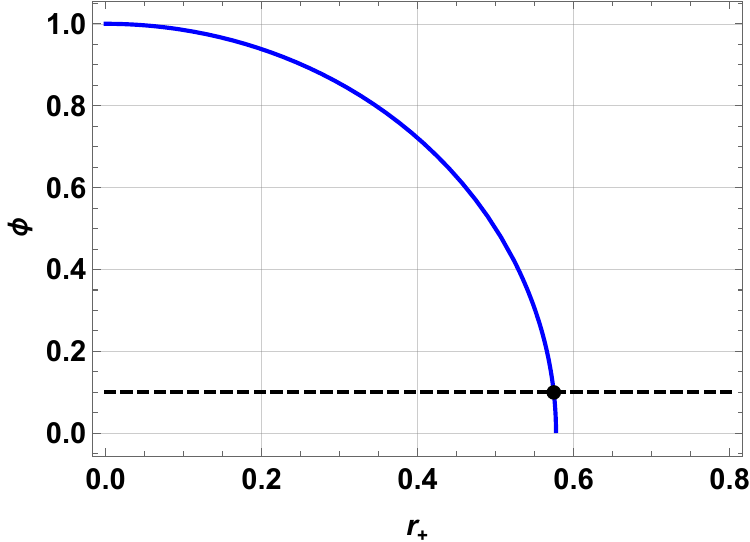}
			\caption{}
			\label{4c}
		\end{subfigure}
		
		\caption{RN AdS black hole : Figure (a) represents $F$ vs $r_+$ plot for $\phi=0.1$. The blue dot represents $r_+=0.99498$, the free energy becomes zero. This is the exact point at which the Hawking-Page transition takes place. Figure (b) represents $C$ vs $r_+$ plot for $\phi=0.1$. The black dot represents at $r_+=0.57445$, the specific heat diverges. This is the exact point at which the Davies-type transition takes place. In Figure (b) spinodal curve is plotted for different values of $\phi$. 
		}
		\label{4}
	\end{figure}

	The equation of motion of the RN-AdS black hole  in 4-dimension is :
	$$ds^2=-f(r)dt^2+\frac{1}{f(r)}dr^2+r^2(d\theta^2+sin^2\theta d\phi^2)$$
	where the metric function $f(r)$ is given by :
	\begin{equation}
	f(r)=1-\frac{2M}{r}+\frac{Q^2}{r^2}+\frac{r^2}{l^2}
	\end{equation}
	here $M$ and $Q$  is the mass and  charge of the black hole respectively and $l$ is the Ads radius.
	At the event horizon radius $r=r_+$ the metric function is set to zero and the mass is obtained as 
	\begin{equation}
		M=\frac{l^2 Q^2+l^2 r_+^2+r_+^4}{2 l^2 r_+}
		\label{rnmass}
	\end{equation}
	In grand canonical ensemble, the charge is converted to its conjugate electric potential as follows: 
	\begin{equation}
		\phi=\frac{\partial M}{\partial Q}=\frac{q}{r_+}
	\end{equation}
	or 
	\begin{equation}
		Q= r _+ \phi
	\end{equation}
	Substituting the value of $Q$ in the expression of mass in eq.(\ref{rnmass},) we get the new mass in grand canonical ensemble  as 
	\begin{equation}
		\tilde{M}=M-Q\phi=\frac{1}{2} r_+ \left(r_+^2-\phi ^2+1\right)
	\end{equation}
	where we have set $l=1$ for simplicity.
	The temperature and the entropy of the black hole is calculated as :
	\begin{equation}
		T=\frac{3 r_+^2-\phi ^2+1}{4 \pi  r_+}
	\end{equation}
	and 
	\begin{equation}
	S=\pi r_{+}^2
	\end{equation}
	The free energy of RN AdS black hole in the grand canonical ensemble is : 
	\begin{equation}
		F=M-T S=-\frac{1}{4} r_+ \left(r_+^2+\phi ^2-1\right)
	\end{equation}
	The plot of free energy is shown in FIG.(\ref{4a}.) The figure shows that the free energy becomes zero at $r_+=0.99498$ for $\phi=0.1$. This point is identified as the Hawking-Page transition point
	he specific heat at constant chemical potential is obtained as :
	\begin{equation}
		C_\phi=T\frac{\partial S}{\partial T}=\frac{2 \pi  r_+^2 \left(3 r_+^2-\phi ^2+1\right)}{3 r_+^2+\phi ^2-1}
	\end{equation}
	The specific heat for $\phi=0.1$ is shown in FIG.(\ref{4b}). The black dot represents the Davies point where the specific heat become zero and it is located at $r_+=0.57445$. The specific heat diverges for $\phi=\sqrt{1-3 r_+^2}$ which is shown in FIG.(\ref{4c}). Each point on the curve represents Davies point.\\
	
	.
	
\begin{figure}[h]	
		\centering
		\includegraphics[width=0.4\linewidth]{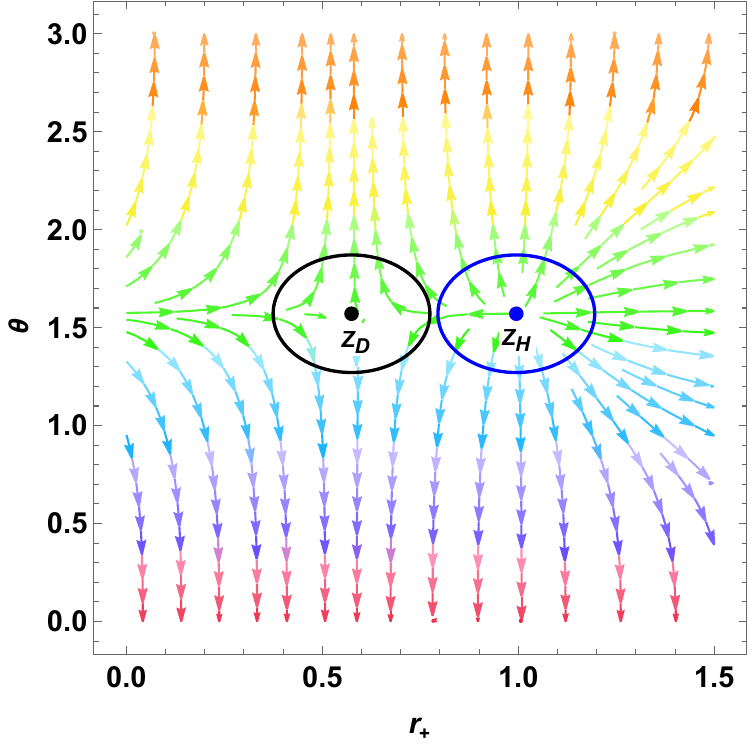}
		\caption{ RN AdS black hole : Vector plot of a portion of the field ($n^1,n^2$). $Z_D$ and $Z_H$ represent the Davies point and Hawking-Page phase transition point respectively.
		}
		\label{5}
	\end{figure}

	Now to study the thermodynamic topology we define the vector $\phi=(\phi^s,\phi^\theta)$ with field components
	\begin{equation}
		\phi^s=\frac{1}{r_+}\frac{\partial F^2}{\partial r_+}=\frac{1}{8} \left(4 r_+^2 \left(\phi ^2-1\right)+3 r_+^4+\left(\phi ^2-1\right)^2\right)
	\end{equation}
	and 
	\begin{equation}
		\phi^\theta=-\cot\theta \csc \theta.
	\end{equation}
	The normalized vector field components are given as
	\begin{equation}
		\begin{cases}
			n^1=\frac{4 r_+^2 \left(\phi ^2-1\right)+3 r_+^4+\left(\phi ^2-1\right)^2}{8 \sqrt{\csc ^4(\theta )-\csc ^2(\theta )+\frac{1}{64} \left(r_+^2+\phi ^2-1\right){}^2 \left(3 r_+^2+\phi ^2-1\right){}^2}}\\
			n^2=-\frac{\cot (\theta ) \csc (\theta )}{\sqrt{\csc ^4(\theta )-\csc ^2(\theta )+\frac{1}{64} \left(r_+^2+\phi ^2-1\right){}^2 \left(3 r_+^2+\phi ^2-1\right){}^2}}
		\end{cases}
	\end{equation}
	and
	The normalized vector is represented in FIG.(\ref{5}) for $\phi=0.1$.\\
	
\begin{figure}[h]	
		\centering
		\begin{subfigure}{0.40\textwidth}
			\includegraphics[width=\linewidth]{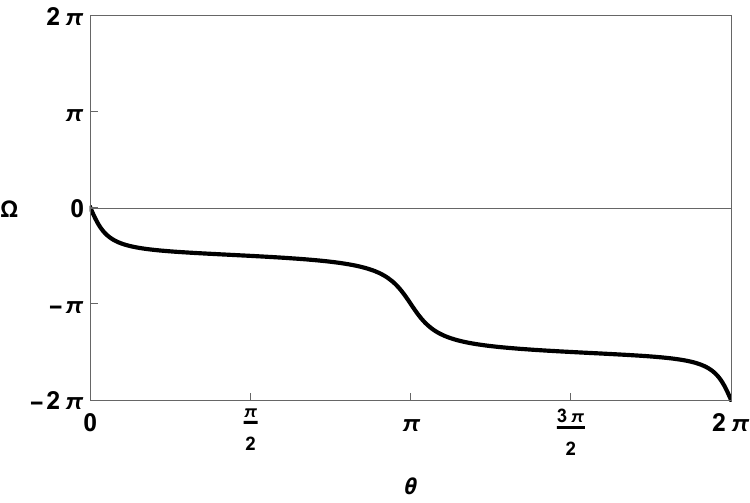}
			\caption{}
			\label{6a}
		\end{subfigure}
		\hspace{0.5cm}
		\begin{subfigure}{0.40\textwidth}
			\includegraphics[width=\linewidth]{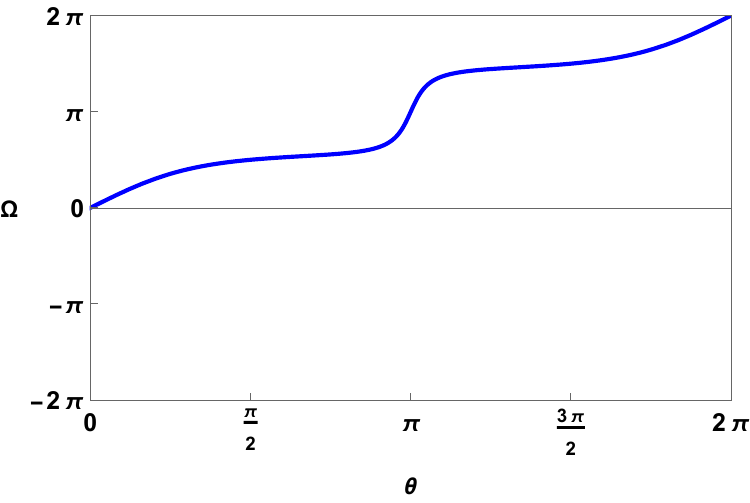}
			\caption{}
			\label{6b}
		\end{subfigure}
		
		\caption{ RN AdS black hole : FIG. (a) represents topological number calculation for Davies point around the black contour in FiG.(\ref{5}) and FIG. (b) represents the topological number calculation for Hawking-Page phase transition point around the blue contour in FIG.(\ref{5}). We have considered $\phi=0.1$.
		}
		\label{6}
	\end{figure}

	The critical points are calculated by setting $\theta=\pi/2$ in $n^1$ and equating it to zero. We find two real and positive critical (or zero) points located at $(r_+,\theta)=(0.57445,\pi/2),(0.99498,\pi/2)$ which are respectively named as $Z_D$ and $Z_H$. These points exactly match with the Davies points and the Hawking-Page phase transition point. 
	We calculate the topological charge associated with each critical point by encircling contours around them as shown in FIG.(\ref{6}). For the critical point $Z_D$ (which corresponds to the Davies point) the topological charge is $-1$ (FIG.(\ref{6a}))and for the critical point $Z_H$ (which corresponds to the Hawking-Page phase transition point) the topological charge is $+1$(FIG.(\ref{6a})).
	
	\section{Kerr Ads Black hole}
\begin{figure}[h!]	
		\centering
		\begin{subfigure}{0.45\textwidth}
			\includegraphics[width=\linewidth]{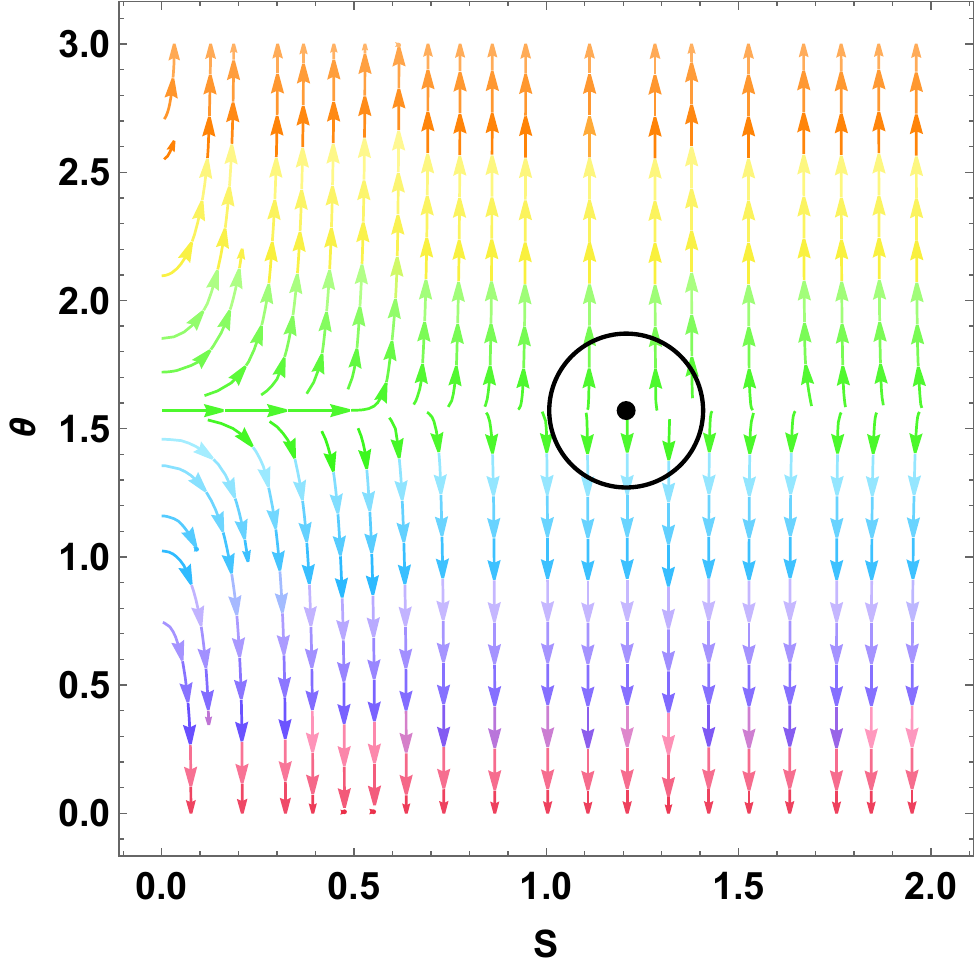}
			\caption{}
			\label{8a}
		\end{subfigure}
		\begin{subfigure}{0.45\textwidth}
			\includegraphics[width=\linewidth]{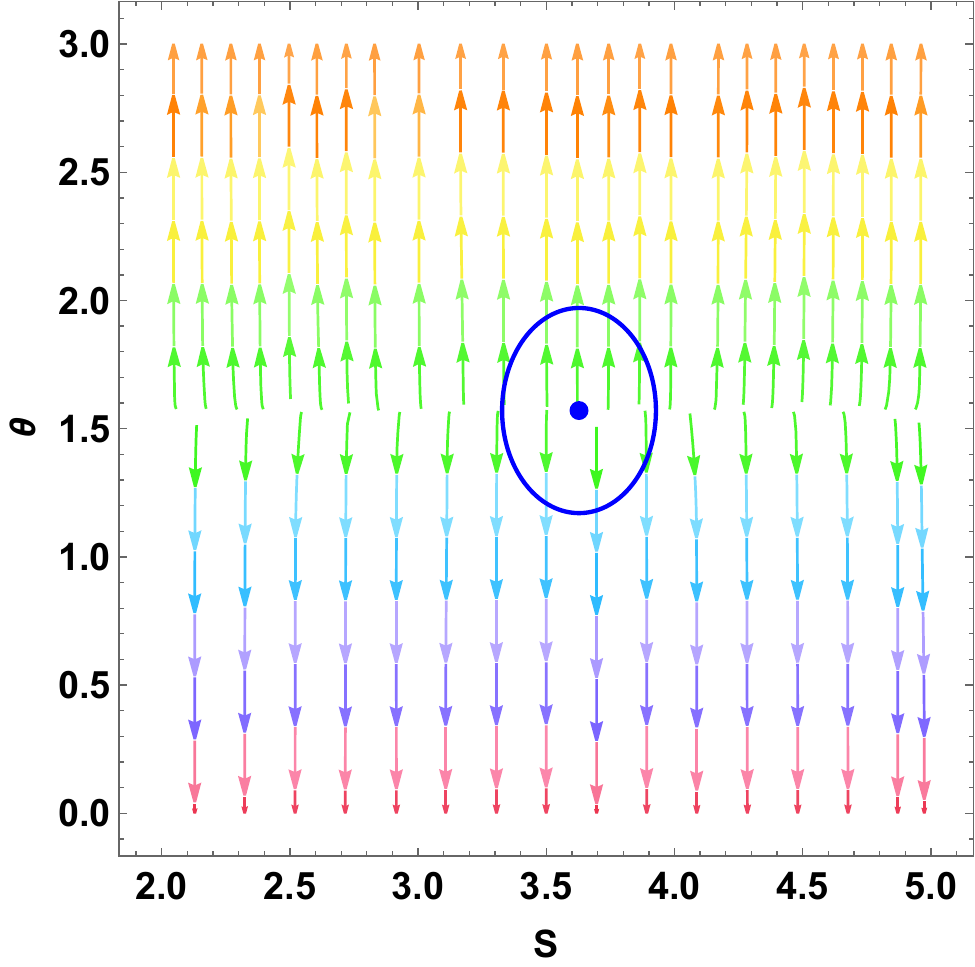}
			\caption{}
			\label{8b}
		\end{subfigure}
		
		\caption{Kerr AdS black hole:  Figure \ref{8a} is a Vector plot of a portion of the vector field ($n^1,n^2$). $Z_D$ and $Z_H$ represent the Davies point and Hawking-Page phase transition point respectively. In Figure.\ref{8b} red and blue lines shows that the topological charge of $Z_D$ is $-1$ and that of $Z_H$ is $1$ respectively.
		}
		\label{8}
	\end{figure}

	The metric for Kerr-Ads black hole  in Boyer-Lindquist coordinate  is given by :
		\begin{equation}
		ds^2=-\frac{\Delta}{\rho^2}\left [ dt-\frac{a sin^2\theta d\varphi}{\Xi} \right ]^2+\frac{\rho^2}{\Delta} dr^2 +\frac{\rho^2}{\Sigma} d\theta^2+\frac{\Xi sin^2 \theta} {\rho^2}\left[a dt-\frac{r^2+a^2}{\Xi} d\varphi\right]^2
		\label{rcmetric}
	\end{equation}
	where 
	$$\Delta=(r^2+a^2)\left(1+\frac{r^2}{l^2}\right)-2m r$$
	$$\Sigma=1-\frac{a^2}{l^2} cos^2\theta \hspace{0.5cm},\hspace{0.5cm} \rho^2=r^2++a^2cos^2\theta \hspace{0.5cm},\hspace{0.5cm} \Xi=1-\frac{a^2}{l^2}$$
	Here $m$ is the mass parameter, $a$ is the angular momentum per unit mass and $l$ is the AdS length. The expressions for ADM  mass $M$ and the angular momentum  $J$ are :
	\begin{equation}
		M=\frac{m}{\Xi^2} \hspace{0.5cm},\hspace{0.5cm} 	J=\frac{a m}{\Xi^2} 
		\label{xi}
	\end{equation}
	Solving $\Delta_r(r=r_+)=0$ we can write mass parameter $m$ as :
	\begin{equation}
	m=\frac{\left(a^2+r_+^2\right) \left(\frac{r_+^2}{l^2}+1\right)}{2 r_+}
	\end{equation}
	The entropy of the black hole is given by :
	\begin{equation}
	S=\frac{\pi (r_{+}^2+a^2)}{\Xi}
	\label{entropy}
	\end{equation}
	Using equation eq.(\ref{xi}) and eq.(\ref{entropy}), the mass of Kerr Ads black hole in terms of entropy $S$ can be written as :   
	\begin{equation}
		M=\frac{\sqrt{4 \pi ^3 J^2 S+4 \pi ^4 J^2+S^4+2 \pi  S^3+\pi ^2 S^2}}{2 \pi ^{3/2} \sqrt{S}}
		\label{mass kerr}
	\end{equation}
	\begin{figure}[h!]	
	\centering
	\begin{subfigure}{0.40\textwidth}
		\includegraphics[width=\linewidth]{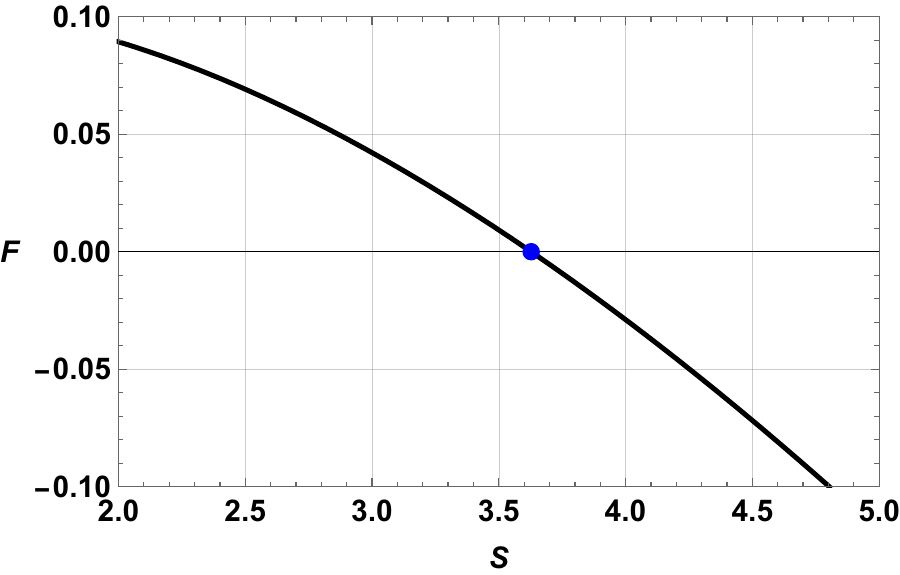}
		\caption{}
		\label{7a}
	\end{subfigure}
	\hspace{0.5cm}
	\begin{subfigure}{0.40\textwidth}
		\includegraphics[width=\linewidth]{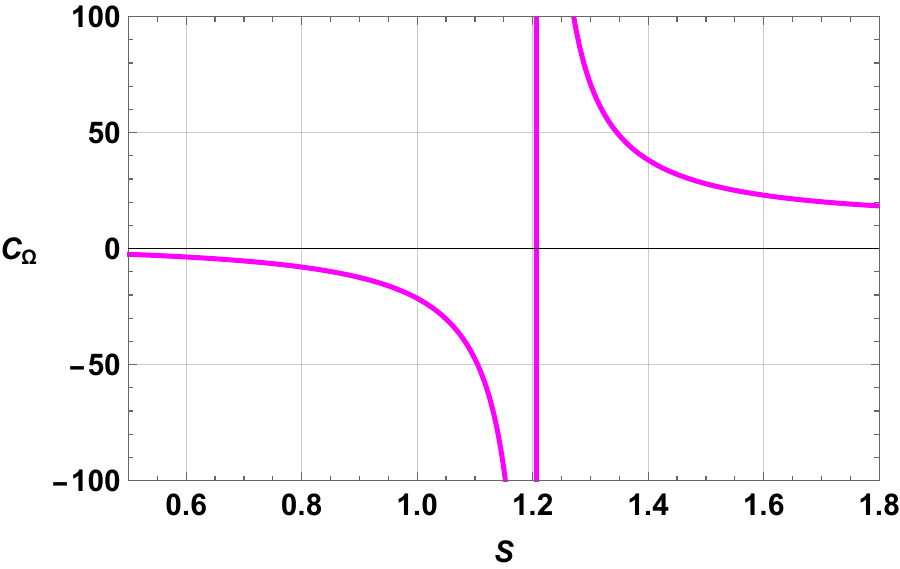}
		\caption{}
		\label{7b}
	\end{subfigure}
	\begin{subfigure}{0.40\textwidth}
		\includegraphics[width=\linewidth]{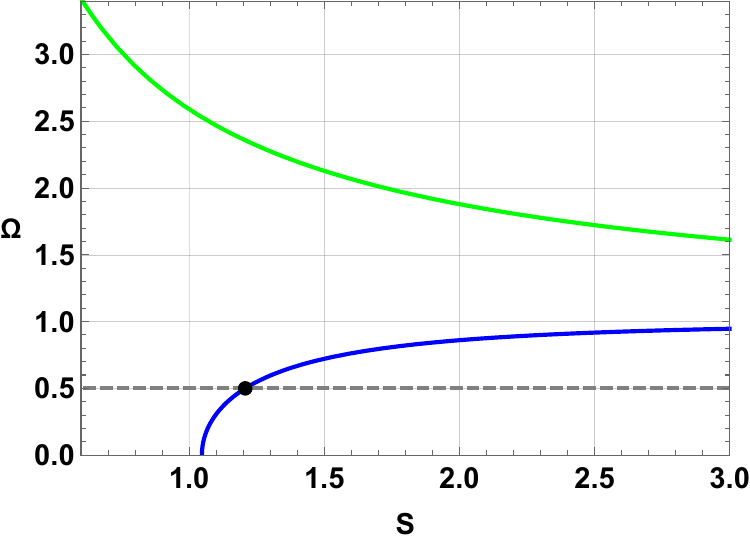}
		\caption{}
		\label{7c}
	\end{subfigure}
	
	\caption{Kerr AdS black hole: FIG. (a) represents $F$ vs $S$ plot for $\Omega=0.5$. The blue dot represents $S=3.62759$, the free energy becomes zero. This is the exact point at which the Hawking-Page transition takes place.  FIG.(b) represents $C$ vs $S$ plot for $\Omega=0.5$. The black dot represents at $S=1.20764$, the specific heat diverges. This is the exact point at which the Davies-type transition takes place. In  FIG.(c) spinodal curve is plotted for different values of $\Omega$.
	}
	\label{7}
\end{figure}
	The angular frequency $\Omega$ is found to be 
	\begin{equation}
		\Omega=\frac{8 \pi ^3 J S+8 \pi ^4 J}{4 \pi ^{3/2} \sqrt{S} \sqrt{4 \pi ^3 J^2 S+4 \pi ^4 J^2+S^4+2 \pi  S^3+\pi ^2 S^2}}
		\label{omega}
	\end{equation}
	Solving eq.\ref{omega}, we get the expression for $J$ as follows:
	\begin{equation}
		J=\frac{S^{3/2} \sqrt{S+\pi } \Omega }{\pi ^{3/2} \sqrt{-4 S \Omega ^2+4 S+4 \pi }}
		\label{j}
	\end{equation}
	Using eq.\ref{mass kerr} and eq.\ref{j}, the new mass in the grand canonical ensemble is found to be :
	\begin{equation}
		\tilde{M}=M-J \Omega= \frac{S^{3/2} (S+\pi )^2}{2 \pi ^{3/2} \sqrt{\frac{S^2 (S+\pi )^3}{-S \Omega ^2+S+\pi }}}
	\end{equation}
	The temperature is given by
	\begin{equation}
		T=\frac{d M}{d S}=\frac{\sqrt{\frac{S^2 (S+\pi )^3}{-S \Omega ^2+S+\pi }} \left(-3 S^2 \left(\Omega ^2-1\right)-2 \pi  S \left(\Omega ^2-2\right)+\pi ^2\right)}{4 \pi ^{3/2} S^{3/2} (S+\pi )^2}
		\label{temp kerr}
	\end{equation}
	Using eq.\ref{mass kerr} and eq.\ref{temp kerr}, We calculate the free energy as :
	\begin{equation}
		F=M-TS=\frac{\sqrt{\frac{S^2 (S+\pi )^3}{-S \Omega ^2+S+\pi }} \left(S^2 \left(\Omega ^2-1\right)+\pi ^2\right)}{4 \pi ^{3/2} \sqrt{S} (S+\pi )^2}
		\label{free kerr}
	\end{equation}
	From Figure.\ref{7a}, the exact point at which the Hawking-Page transition takes place is found to be $S=3.62759$ \\ 
	The specific heat of the system is:
	\begin{equation}
		C=T\frac{\partial S}{\partial T}=\frac{2 s (s+\pi ) \left(s \left(\omega ^2-1\right)-\pi \right) \left(3 s^2 \left(\omega ^2-1\right)+2 \pi  s \left(\omega ^2-2\right)-\pi ^2\right)}{3 s^4 \left(\omega ^2-1\right)^2+4 \pi  s^3 \left(\omega ^4-3 \omega ^2+2\right)-6 \pi ^2 s^2 \left(\omega ^2-1\right)-\pi ^4}
	\end{equation}
	The Davies point is found to be $S=1.20765$ in figure.\ref{7b}.\\
	Now applying eq.\ref{free kerr} in eq. \ref{newfield}, we get the component as :
	\begin{equation}
		\phi^S=\frac{\left(s^2 \left(\omega ^2-1\right)+\pi ^2\right) \left(-3 s^4 \left(\omega ^2-1\right)^2-4 \pi  s^3 \left(\omega ^4-3 \omega ^2+2\right)+6 \pi ^2 s^2 \left(\omega ^2-1\right)+\pi ^4\right)}{16 \pi ^3 s (s+\pi )^2 \left(-s \omega ^2+s+\pi \right)^2}
	\end{equation}
	and 
	\begin{equation}
		\phi^\theta=-\cot (\theta ) \csc (\theta )
	\end{equation}
\vspace{0.5cm}\\
	Using the components the unit vectors $\left(n^1,n^2\right)$ are calculated. 
	We depict two different segments of the vector field ($n^1,n^2$) in Figure.\ref{8}. In Figure.\ref{8a}, one critical point is found at $Z_D=1.20764$ which corresponds to Davies type phase transition, while the other appears at $Z_H=3.6276$, which represents the Hawking-Page phase transition point.\\
	\vspace{1cm}
	\begin{figure}[h!]	
	\centering
	\begin{subfigure}{0.45\textwidth}
		\includegraphics[width=\linewidth]{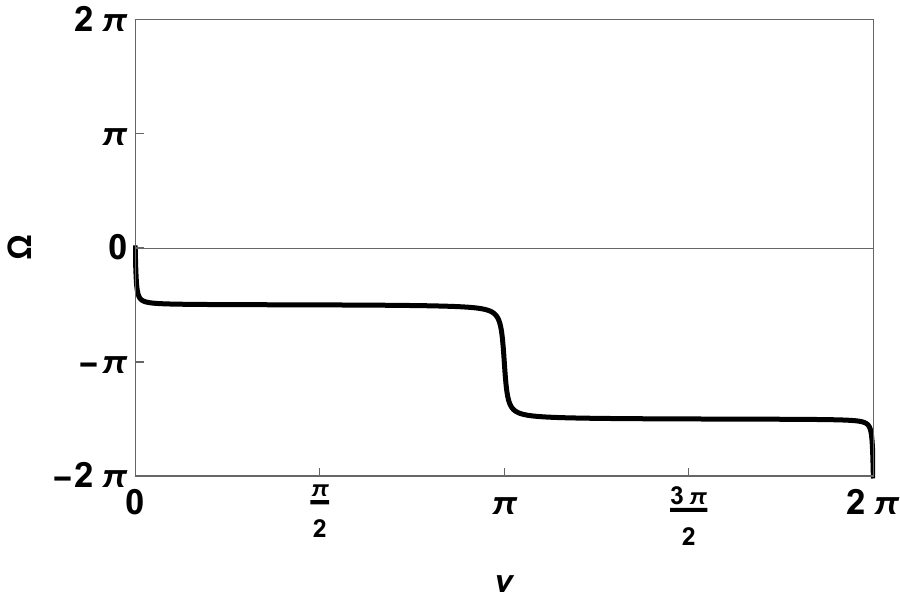}
		\caption{}
		\label{9a}
	\end{subfigure}
	\begin{subfigure}{0.45\textwidth}
		\includegraphics[width=\linewidth]{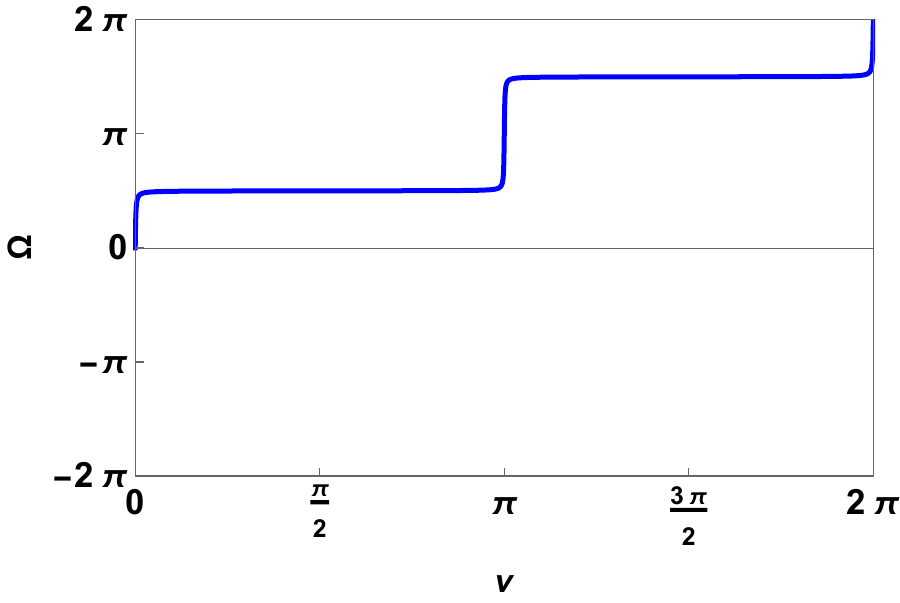}
		\caption{}
		\label{9b}
	\end{subfigure}
	
	\caption{Here black and blue solid line shows that the topological charge of $Z_D$ is $-1$ and that of $Z_H$ is $1$ repectively.
	}
	\label{9}
\end{figure}

	Here also the topological charge corresponding to the Davies point is determined to be $-1$, while for the Hawking-Page phase transition, it is $1$, as illustrated by the black and blue solid line respectively in FIG.(\ref{9}).

	\newpage
	\section{Conclusion}
	We have proposed a common vector field, denoted as $\phi$, within which the thermodynamic topology of the Davies-type and Hawking-Page phase transitions can be studied. The critical points, or zero points, of this vector field precisely coincide with the Davies point and the Hawking-Page phase transition point. Crucially, we can differentiate between these two points by simply calculating the topological charge associated with them. Specifically, the topological charge for the critical point corresponding to the Davies-type phase transition is found to be $-1$, while for the Hawking-Page phase transition point, it is found to be $+1$. Furthermore, we illustrated this concept using three simple systems as examples: the Schwarzschild AdS black hole, the Reissner-Nordström AdS black hole in the grand canonical ensemble, and the Kerr AdS black holes in the grand canonical ensemble. The critical points obtained for these three systems exactly match the Davies-type and Hawking-Page phase transition points. Additionally, the associated topological charge is again $-1$ for the Davies point and $+1$ for the Hawking-Page phase transition point, confirming the validity of our approach.\\
\section{Acknowledgments}
BH would like to thank DST-INSPIRE, Ministry of Science and Technology fellowship program, Govt. of India for awarding the DST/INSPIRE Fellowship[IF220255] for financial support. 	
	
	\bibliographystyle{apsrev}
	
\end{document}